\documentclass[a4paper,11pt]{article}
\usepackage{pos}

\title{Doubly charmed tetraquark $T_{cc}^+$ in (2+1)-flavor QCD near physical point}

\author[a,b]{Yan Lyu}
\author*[c,b]{Sinya Aoki}
\author[b]{Takumi Doi}
\author[b]{Tetsuo Hatsuda}
\author[d]{Yoichi Ikeda}
\author[a,c]{Jie Meng}

\affiliation[a]{State Key Laboratory of Nuclear Physics and Technology, School of Physics, Peking University, Beijing 100871, China
 }
\affiliation[b]{Interdisciplinary Theoretical and Mathematical Sciences Program (iTHEMS), RIKEN,  Wako 351-0198, Japan}
\affiliation[c]{Center for Gravitational Physics and Quantum Information, Yukawa Institute for Theoretical Physics, Kyoto University, 
Kyoto 606-8502, Japan}
\affiliation[d]{Center for Infectious Disease Education and Research, Osaka University, Suita 565-0871, Japan}

\emailAdd{helvetia@pku.edu.cn}
\emailAdd{saoki@yukawa.kyoto-u.ac.jp}
\emailAdd{doi@ribf.riken.jp}
\emailAdd{thatsuda@riken.jp}
\emailAdd{yikeda@cider.osaka-u.ac.jp}
\emailAdd{mengj@pku.edu.cn}

\abstract{We study the doubly charmed tetraquark state $T_{cc}^+$ by the HAL QCD method applied to the $D^*D$ system in $(2+1)$ flavor lattice QCD at nearly physical pion mass,  $m_\pi= 146$ MeV.
We obtain the attractive potential at all distances in the $S$-wave of the isoscalar $D^* D$ system, whose long distance behavior is well described by the 
two-pion exchange (TPE), and  it generates a virtual pole near $D^* D$ threshold
with a pole position $E_{\rm pole} = -59 (^{+53}_{-99}) (^{+2}_{-67})$ keV and an inverse scattering length $1/a_0=0.05(5)(^{+2}_{-2})$ fm$^{-1}$.   
The virtual pole turns into  a loosely bound state pole if the pion mass in the TPE potential is extrapolated  to the physical value, $m_\pi =135$ MeV. 
The potential at the physical pion mass is shown to give  a semi-quantitative description of the $D^0 D^0\pi^+$ mass spectrum at the LHCb. 
}

\FullConference{The 40th International Symposium on Lattice Field Theory (Lattice 2023)\\
July 31st - August 4th, 2023\\
Fermi National Accelerator Laboratory\\}

\begin{document}
\begin{flushright}
YITP-24-06, RIKEN-iTHEMS-Report-24
\end{flushright}
\maketitle

\section{Introduction}
A heavy tetraquark state which at least includes two heavy quarks and two light anti-quarks is one of the clearest exotic hadrons,
since it never mixes with an ordinary meson made of quark and anti-quark.
A candidate of a doubly charmed tetraquark state $T^+_{cc}$ was experimentally observed by the LHCb collaboration\cite{LHCb:2021vvq,LHCb:2021auc},
who reported a narrow peak with $I(J^P) =0(1^+)$
in the $D^0D^0\pi^+$ mass spectrum at 360 keV below $D^{*+}D^0$ threshold.

\begin{figure}[htb]
\centering
  \includegraphics[angle=0, width=0.45\textwidth]{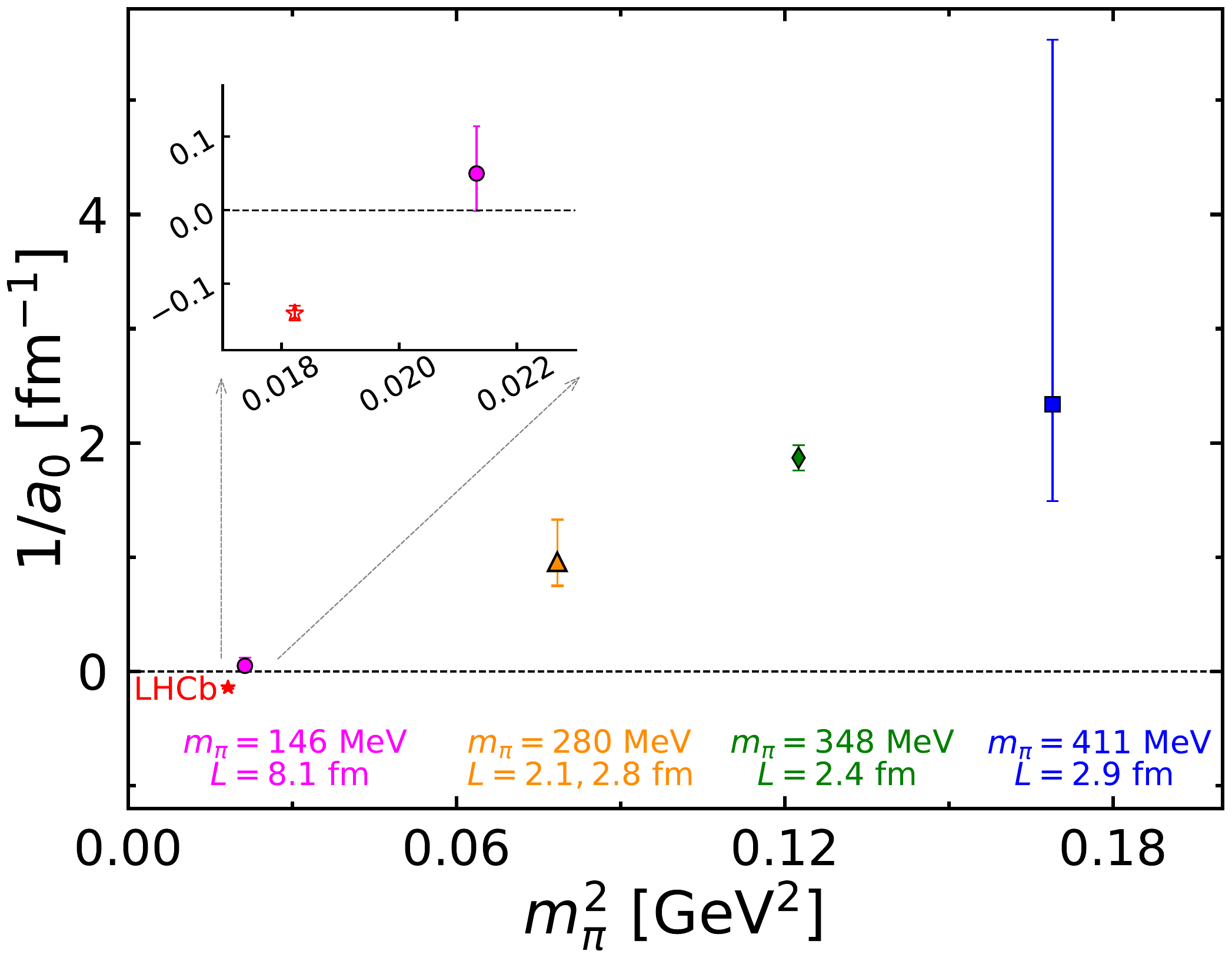}
  \hskip 0.5cm
    \includegraphics[angle=0, width=0.45\textwidth]{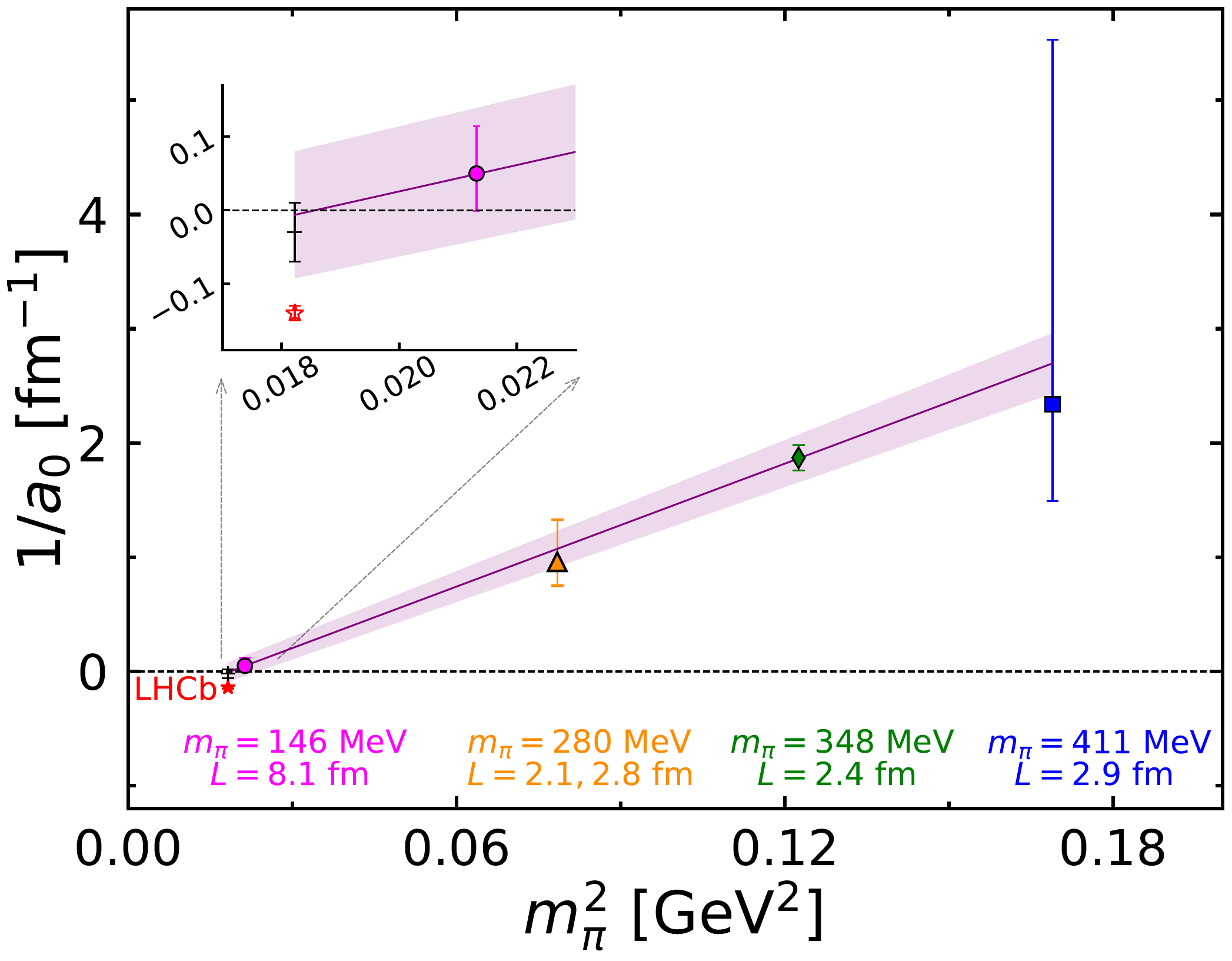}
 \caption{(Left) The inverse scattering length $1/a_0$ for the $S$-wave $D^*D$ system with $I=0$ as a function of $m_\pi^2$, obtained from lattice QCD calculations by Refs.~\cite{Ikeda:2013vwa} (blue square),  \cite{Chen:2022vpo}(green diamond), \cite{Padmanath:2022cvl}(yellow triangle), and \cite{Lyu:2023xro}(magenta circle), together with the real part of the experimental value by LHCb (red star)\cite{LHCb:2021vvq}.
 (Right) The chiral extrapolation of $1/a_0$ linear in $m_\pi^2$ using lattice data.  The black plus shows the value obtained from the potential modified to $m_\pi=135$ MeV in the TPE contribution.
  }
 \label{fig:a0}
\end{figure}

In the theoretical side, the first-principle lattice QCD calculations exhibit a strong quark mass dependence of the  $D^{*+}D^0$ interaction
in this channel, as shown in Fig.~\ref{fig:a0} (Left), where the inverse scattering length $1/a_0$ for the $S$-wave $D^* D$ system with $I=0$ is plotted as a function of $m_\pi^2$.  As the inverse scattering length decreases and becomes very small toward the physical pion mass $m_\pi^2\simeq 0.018$ GeV$^2$,  the lattice calculation should be performed near physical pion mass to make a reliable comparison between the theoretical prediction and the experimental data. Such a study has been performed in Ref.~\cite{Lyu:2023xro}(magenta circle), whose results are reported here.

\section{Methodology}
\subsection{HAL QCD method}
In the HAL QCD method\cite{Ishii:2006ec}, we consider the equal-time Nambu-Bethe-Salpeter (NBS) wave function for the $D^* D$ system given by
\begin{equation}
\Psi_{W}^{D^* D} ({\bf r}) e^{-W t} :={1\over {\sqrt{Z_{D^*}}}}{1\over {\sqrt{Z_D}}} \sum_{\bf x} \langle 0 \vert D^*({\bf x} +{\bf r},t) D({\bf x}, t) \vert D^* D; W \rangle,  
\end{equation}
where $\vert 0\rangle$ and $\vert D^* D; W \rangle$ are the vacuum and the $D^* D$ eigenstate with the center of mass energy $W$, respectively, and
$Z_{D^*}$ and   $Z_D$ are wave function renormalization factors for single $D^*$ and $D$ mesons, respectively. 
Meson operators in $\Psi_W$ are given by $ D (x) = \bar q(x) \gamma_5 c(x)$ and $ D^* (x) = \bar q(x) \gamma_i c(x)$, where
$\bar q(x)= \bar u(x)$ or $\bar d(x)$ represents the light anti-quark. 

 One can define a non-local but energy-independent potential from the NBS wave function as
 \begin{equation}
\left({\nabla^2\over 2\mu} + {p_W^2\over 2\mu}\right) \psi_W^{D^* D}(\mathbf{r}) = \int d^3\mathbf{r}^\prime U(\mathbf{r}, \mathbf{r}^\prime)  \psi_W^{D^* D}(\mathbf{r}^\prime),
\end{equation}
where $p_W$ is the magnitude of the relative momentum defined by $W=\sqrt{p_W^2+ m_D^2} + \sqrt{p_W^2+ m_{D^*}^2}$, and $\mu$ is a reduced mass as $1/\mu = 1/m_D + 1/m_{D^*}$.

In practice, the non-locality is treated by the derivative expansion as $\displaystyle U(\mathbf{r}, \mathbf{r}^\prime) =\sum_{k=0}^\infty V^{(k)}(\mathbf{r} ) \nabla^k \delta(\mathbf{r} -\mathbf{r}^\prime)$. For example, at the leading order (LO) of the derivative expansion, 
we obtain
\begin{equation}
V^{(0)}(\mathbf{r};W) = {1\over   \psi_W^{D^* D}(\mathbf{r}) } \left( -H_0 + {p_W^2\over 2\mu}\right) \psi_W^{D^* D}(\mathbf{r}) ,
\quad H_0:=  -{\nabla^2\over 2\mu}, 
\end{equation}
 where the argument $W$ in the LO potential represents a fact that the LO potential is determined from the NBS wave function at the energy $W$.
 While the LO analysis is employed throughout this paper, it is possible to obtain higher order terms in some cases, which improved accuracy
 of results\cite{HALQCD:2018gyl}. 
 
One can extract the NBS wave function from the 4-pt correlation functions, which can be calculated in lattice QCD,
and then extract the potential.  
In this study, we have employed  the time-dependent HAL QCD method\cite{Ishii:2012ssm},
which gives the LO potential as
 \begin{equation}
V^{(0)}(r;t) ={1\over R(\mathbf{r},t)}\left[ {1+ 3\delta^2\over 8\mu} \partial_t^2 -\partial_t -H_0\right] R(\mathbf{r},t), \quad
\delta:={m_{D^*}-m_D\over m_{D^*} + m_D}, 
\end{equation}
where 
 \begin{equation}
R(\mathbf{r},t)e^{-(m_{D^*}+m_D )t} = \sum_{\mathbf{x}}  \langle 0 \vert D^*({\bf x} +{\bf r},t) D({\bf x}, t) \overline{\cal J}(0)\vert 0\rangle,
\end{equation}
$ \overline{\cal J}$ is the wall-type source which creates $D^* D$ states in the $I=0$ and $S$-wave channel, and the
$S$ wave projection of $R(\mathbf{r},t)$ is performed on the lattice\cite{Miyamoto:2019jjc}.
The neglected contribution at $O(\delta^2\partial_t^3)$ in the above extraction is consistent with zero within statistical uncertainties in this study.
A weak $t$-dependence of $V^{(0)}(r;t)$ indicates that  contributions from inelastic states as well as higher order terms in the derivative expansion are well under control. 
 
 \subsection{Lattice setup} 
 \begin{figure}[htb]
\centering
  \includegraphics[angle=0, width=0.7\textwidth]{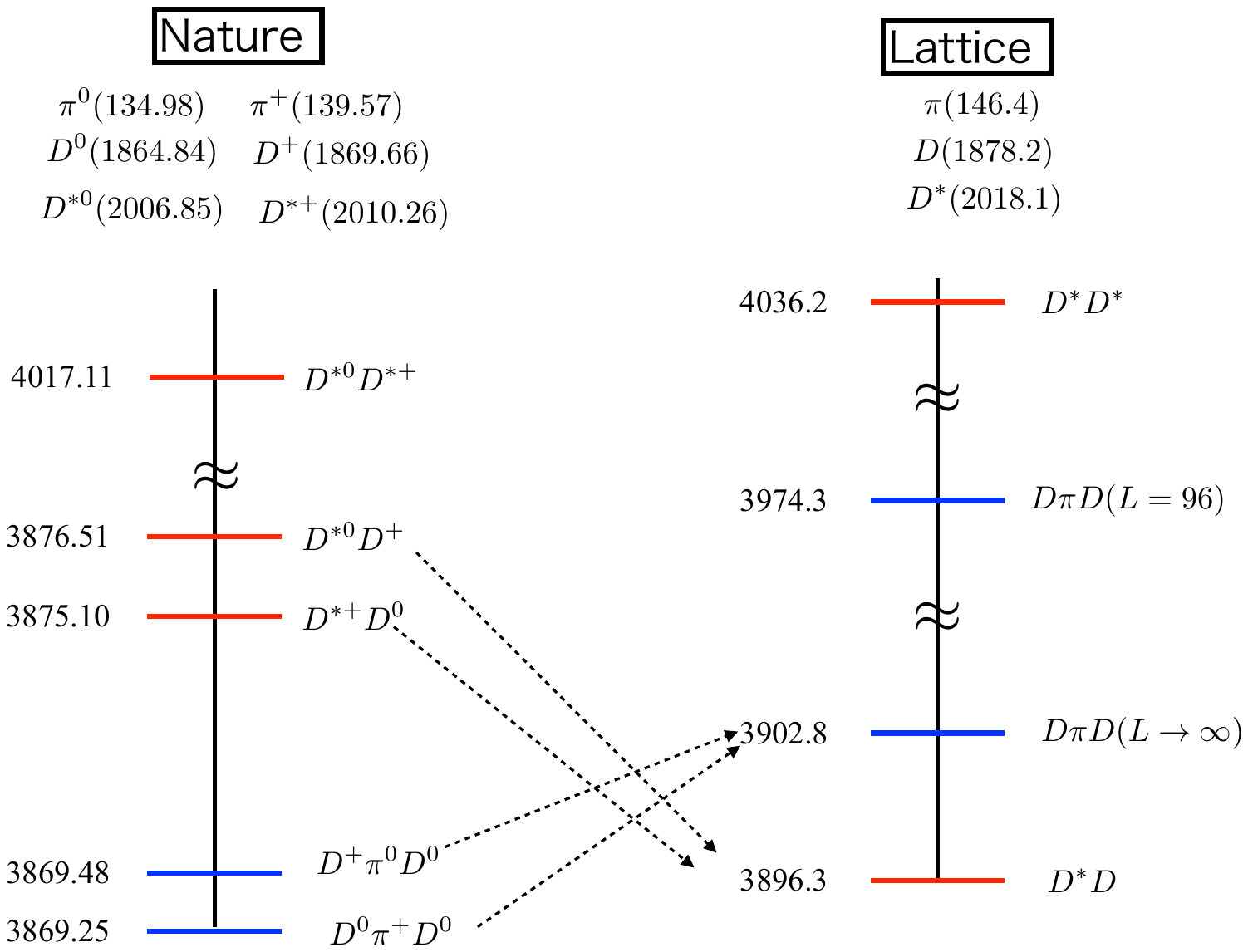}
 \caption{The ordering of thresholds in the $I=0$ $D^*D$ system in Nature (Left) or in our lattice setup (Right).  }
 \label{fig:threshold}
\end{figure}
In this study, we have employed $(2+1)$-flavor gauge configurations on the $96^4$ lattice generated with the Iwasaki gauge action and the nonperturbatively $O(a)$ improved Wilson quark action at nearly physical pion mass $m_\pi=146.4$ MeV with lattice spacing $a= 0.0846$ fm, corresponding to lattice size $L=8.1$ fm\cite{Ishikawa:2015rho}.
The charm quark is treated in the quenched approximation by using the relativistic heavy quark (RHQ) action proposed in Ref.~\cite{Aoki:2001ra}, and
the charm quark mass is taken to give a spin-averaged $1S$ charmonium mass $M_{\rm av}:= ( m_{\eta_c} + 3 m_{J/\Psi})/4 =3096.6(0.3)$ MeV\cite{Namekawa:2017nfw}, which is 0.9\% larger than the experimental value   $M_{\rm av}=3068.5$ MeV.
We have confirmed that  the effect of this mass difference to physical observables is small compared with statistical uncertainties, by performing an extra calculation with a different charm quark mass corresponding to $M_{\rm av}=3051.4(0.3)$ MeV.

In Nature with isospin symmetry breaking, there appear 3-body channels below the threshold of the $D^* D$ system with $I=0$, as shown in Fig.~\ref{fig:threshold} (Left).
In our lattice setup, however, the $D^* D$ becomes the lowest channel in the system, as can be seen in Fig.~\ref{fig:threshold} (Right).

 \section{Numerical results}
  \subsection{$D^*D$ potential and its long distance behavior}
\begin{figure}[htb]
\centering
  \includegraphics[angle=0, width=0.45\textwidth]{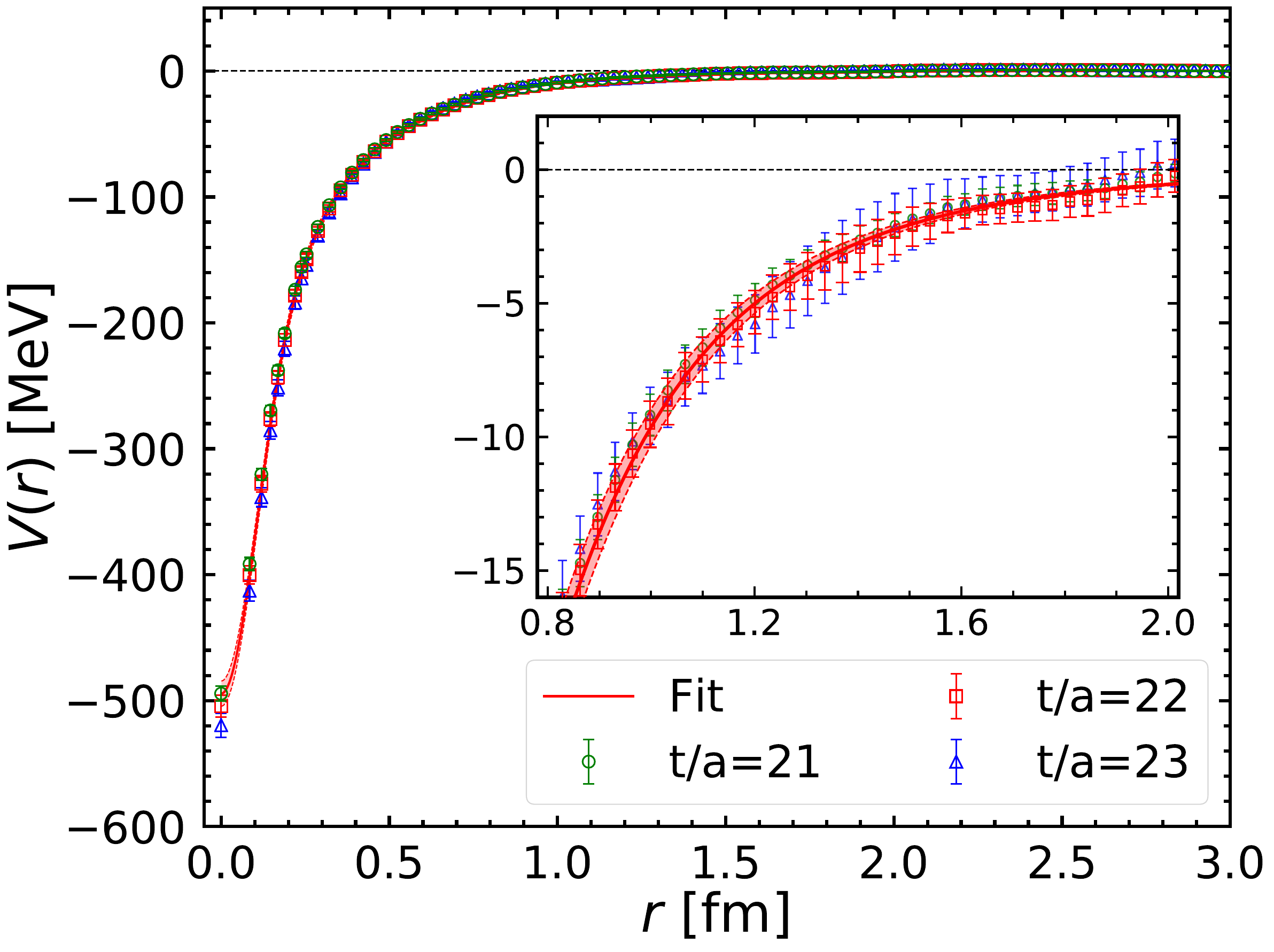}
  \hskip 0.5cm
    \includegraphics[angle=0, width=0.45\textwidth]{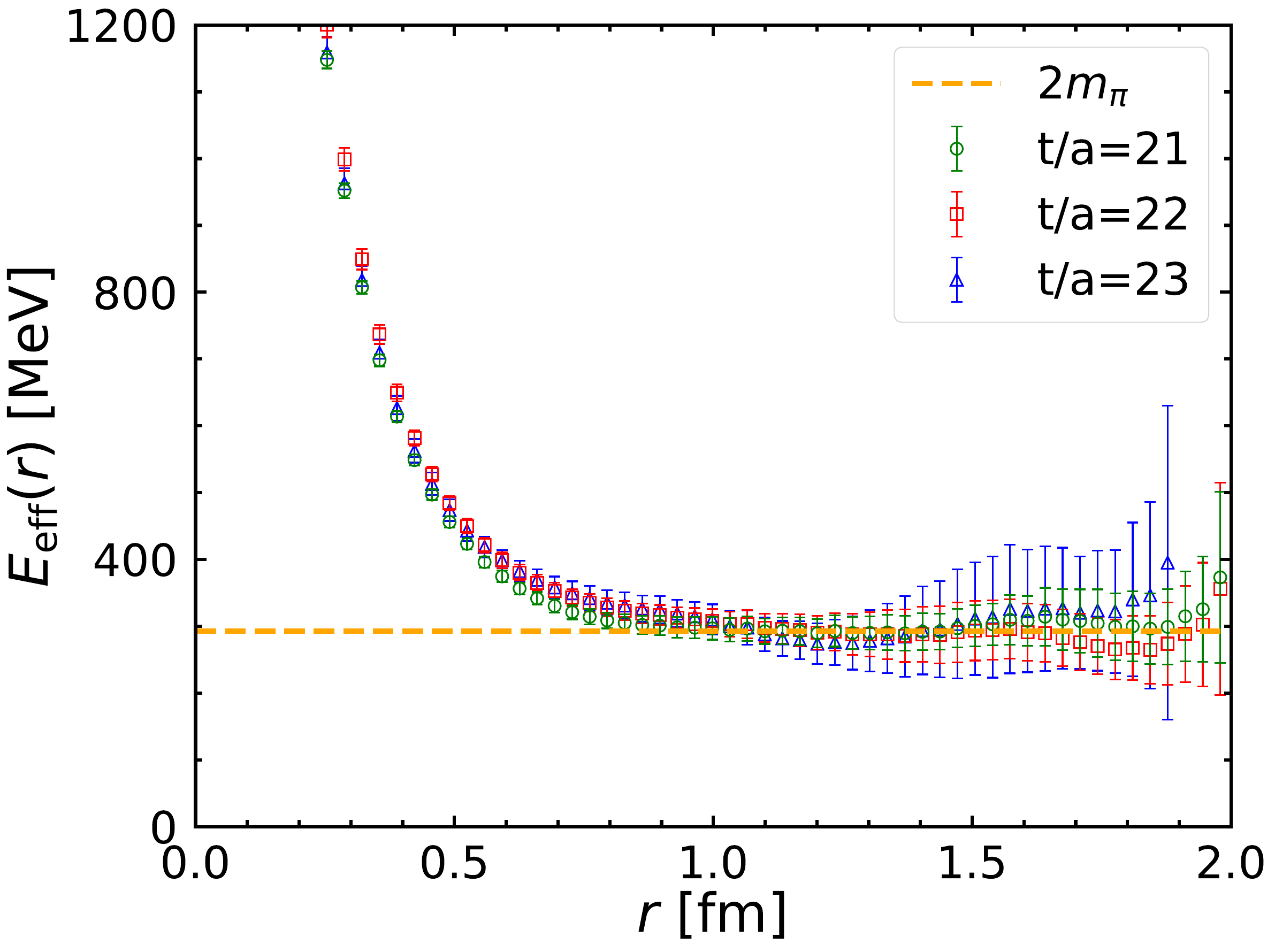}
 \caption{(Left) The LO $D^*D$ potential in the $I=0$ and $S$-wave  channel as a function of the distance $r$.
 (Right) The effective energy in space as a function of the distance $r$ at $t/a=21$ (green circles), 22 (red squares) and 23 (blue triangles), together with the horizontal orange dashed line at $2 m_\pi = 292.8$ MeV. }
 \label{fig:potential}
\end{figure}

Fig.~\ref{fig:potential} (Left) shows the LO $D^*D$ potential $V^{(0)}(r;t)$ in the $I=0$ and $S$-wave  channel for $t/a=21,22,23$, corresponding to $t\simeq 1.9$ fm in physical unit, which is chosen to suppress inelastic states contributions at small $t$ and at the same time to avoid large statistical uncertainties at large $t$. 
We observe a weak $t$ dependence of $V^{(0)}(r;t)$ in Fig.~\ref{fig:potential} (Left), which indicates that both effects are small, and
we take small variations for different $t$ into account as a source of systematic errors.

The LO $D^*D$ potential $V^{(0)}(r;t)$  in the $I(J^P)=0(1^+)$ channel in Fig.~\ref{fig:potential} (Left) shows attractive behaviors at all distances,
whose short-range part may be related to the coupling between antidiquark and $D^*D$ state\cite{Jaffe:2003sg,Lee:2009rt}. 
Similar short-range attraction has been also observed on the lattice for the $B^*B$ system in the  $I(J^P)=0(1^+)$ channel\cite{Bicudo:2015kna,Aoki:2023nzp}.

The long-range part of the attraction for $r > 1$ fm is expected to be explained by one-pion exchange (OPE) between $D^*$ and $D$, which behaves
as $e^{-m r}/r$ with either $m=m_\pi\simeq 146$ MeV\cite{Ohkoda:2012hv} or $m=\sqrt{m_\pi^2-(m_{D^*} - m_D)^2}\simeq 43$ MeV\cite{Li:2012ss}. 
We have found, however, that the longe-range part of the potential cannot be well reproduced by the OPE potential.
Therefore,  motivated by the recent finding\cite{Lyu:2022imf}, we instead  include two-pion exchange (TPE) potential in our fit function as
\begin{eqnarray}
V_{\rm fit} (r;m_\pi) =\sum_{i=1,2} a_i e^{-(r/b_i)^2} + a_3 \left( 1-e^{-(r/b_3)^2}\right)^2 {e^{-2m_\pi r}\over r^2},
\label{eq:Fit}
\end{eqnarray}
which turns out to reproduce data well, as shown by the red band in Fig.~\ref{fig:potential} (Left).
To validate an existence of the TPE contribution in the long-range part of the potential, we calculate the spatial effective energy, defined by
\begin{equation}
E_{\rm eff}(r) = - {1\over r} \ln \left[{V^{(0)}(r;t) r^2 \over a_3}\right],
\end{equation}
which is plotted in Fig.~\ref{fig:potential} (Right).
The effective energy reaches a plateau at $2m_\pi = 292.8$ MeV for  $r \ge 1$ fm,
which indicates that the long-range part of the potential is well described by the TPE,
though the fitted parameter $a_3$ is used here. 

  \subsection{Scattering parameters}
\begin{figure}[htb]
\centering
  \includegraphics[angle=0, width=0.45\textwidth]{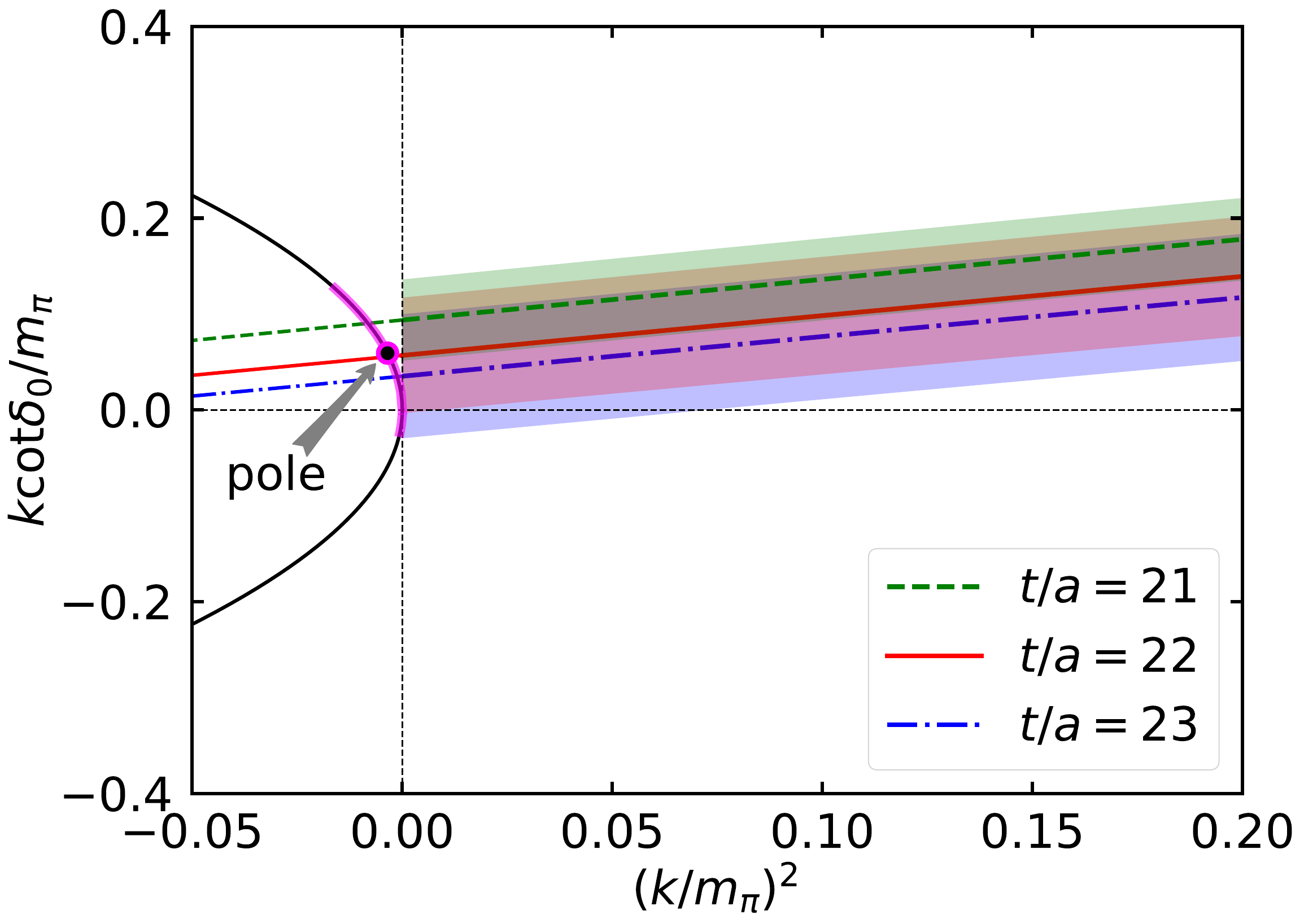}
  \hskip 0.5cm
    \includegraphics[angle=0, width=0.45\textwidth]{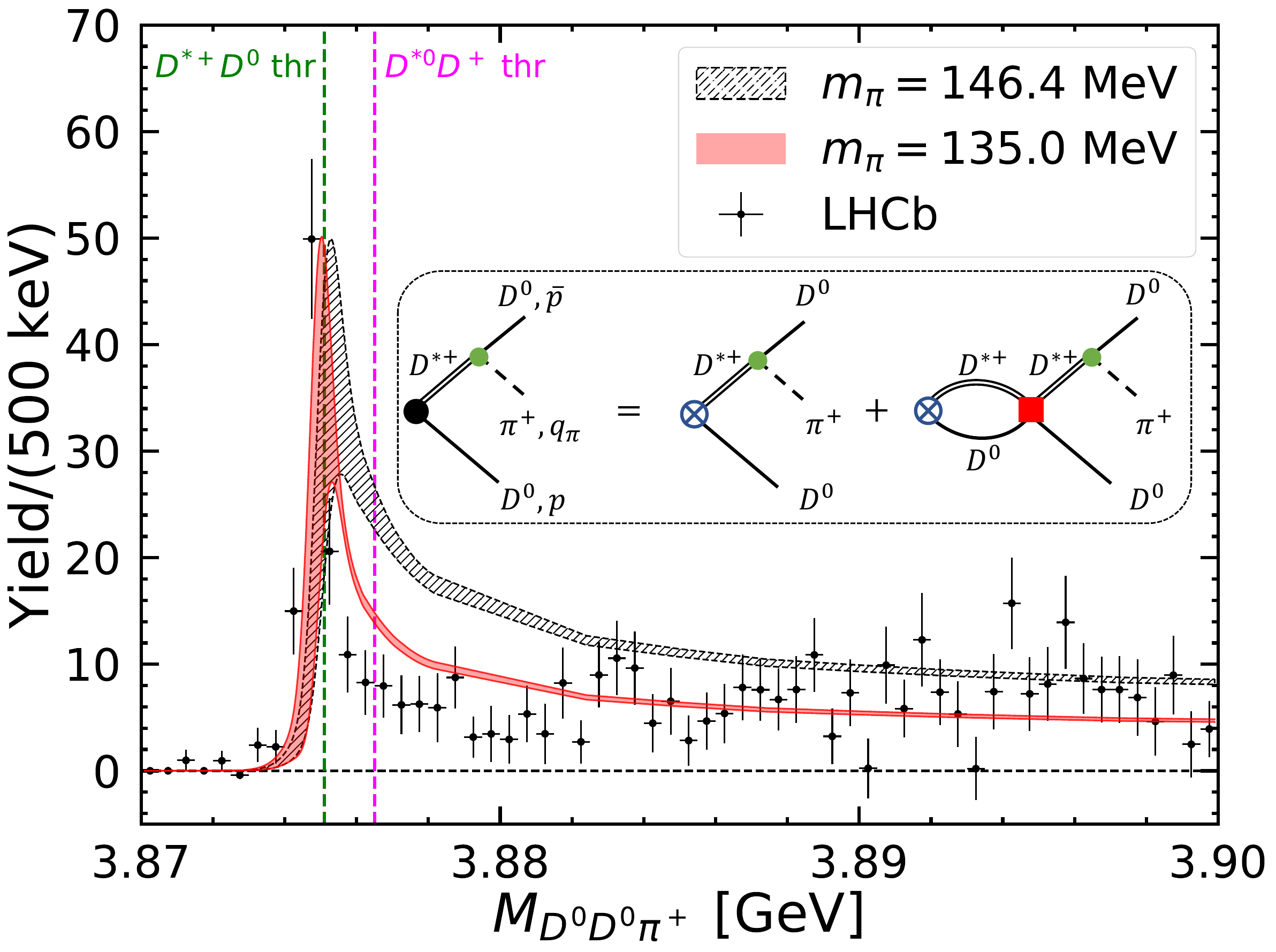}
 \caption{(Left) The $k\cot \delta_0(k)/m_\pi$ for the $S$-wave $D^* D$ scattering in the $I=0$ channel as a function of $k^2/m_\pi^2$,
 calculated from the fitted potential at $t/a=21$ (green dashed line), 22 (red solid line) and 23 (blue dashed-dotted line).
 The intersection between the ERE line and $+\sqrt{ -k^2/m_\pi^2}$ (the black solid line) represents a virtual pole of the scattering amplitude, where the black point indicates its central value at $t/a=22$ while the magenta line denotes statistical and systematic errors combined.
 (Right) The $D^0D^0\pi^+$ mass spectrum predicted with lattice potentials at $m_\pi=146.4$ MeV (black band) and $m_\pi=135$ MeV (red band),
 together with  LHCb data (black points). 
 The insect shows diagrams for contributions to the $D^0D^0\pi^+$ mass spectrum, where
 black filled circle, blue cross circle, green filled circle, and red filled square denote production amplitude $U$, constant vertex $P$, $D^{*+}\to D^0\pi^+$ vertex, and scattering $T$ matrix, respectively, and there exists another diagram with momentum $p$ and $\bar p$ exchanged.
  }
 \label{fig:ERE}
\end{figure}

Using the fitted potential, 
we then calculate the $S$-wave  $D^* D$ scattering phase shift $\delta_0$ in the $I=0$ channel through the Schr\"odinger equation
in the infinite volume with measured value of $m_{D^*}$ and $m_D$ on the lattice.
 Fig.~\ref{fig:ERE} (Left) shows $k\cot \delta_0(k)/m_\pi$ as a function of $k^2/m_\pi^2$ with a relative momentum $k$,
 which is fitted by the effective range expansion (ERE) as
 \begin{equation}
k\cot \delta_0(k) = {1\over a_0} + {1\over 2} r_{\rm eff} k^2 + O(k^4),
\end{equation}
where $a_0$ is the scattering length with the sign convention of high-energy physics and $r_{\rm eff}$ is the effective range.
We obtain $1/a_0 = 0.05(5)(^{+2}_{-2})$ fm$^{-1}$ and $r_{\rm eff}=1.12(3)(^{+3}_{-8})$ fm, the former of which is shown in Fig.~\ref{fig:a0} (Left) by the magenta circle, together with the previous lattice results and the LHCb data,
where numbers in the first parenthesis represent statistical errors while the second ones systematic errors\cite{Lyu:2023xro}. 
The $S$-wave $D^*D$ system  in the $I=0$ channel  appears in the unitary regime ($1/a_0\sim 0$) at $m_\pi= 146$ MeV.  
As seen from the figure, the ERE intersects with $+\sqrt{-k^2/m_\pi^2}$ at negative $k^2$, which leads to one shallow virtual pole 
with $\kappa_ {\rm pole} = -8(8)(^{+3}_{-5})$ MeV and $E_ {\rm pole} = -59(^{+53}_{-99})(^{+2}_{-67})$ keV, 
where $k=i \kappa_ {\rm pole}$ and $E_{\rm pole} = \sqrt{m_{D^*}^2-\kappa_ {\rm pole}^2}+  \sqrt{m_{D}^2-\kappa_ {\rm pole}^2} -(m_{D^*} + m_D)$. 
 
 To estimate effects of a small difference between $m_\pi=146$ MeV and the physical pion mass on the scattering parameters,
 we replace the pion mass  in the TPE potential (the last term in Eq.~\eqref{eq:Fit}) by $m_\pi = 135$ MeV, the physical pion mass without the QED contribution\cite{FlavourLatticeAveragingGroupFLAG:2021npn}, while other parameters ($a_{1,2,3}, b_{1,2,3}$) are kept fixed to values at $t/a=22$.
Employing such a modified potential with physical $m_{D^{*+}}$ and $m_{D^0}$, we obtain one loosely bound state with
$1/a_0= -0.03(4)$ fm$^{-1}$, $r_{\rm eff} =1.12(3)$ fm, $\kappa_{\rm pole} = + 5(8)$ MeV, and $E_{\rm pole} = -45(^{+41}_{-78})$ keV,    
 which indicates an existence of a bound state $T_{cc}^+$ at physical point, though its binding energy is much smaller than
  $E_{\rm pole} = -360(40)(^{+4}_{-0})$ keV reported by LHCb\cite{LHCb:2021vvq}. 
  
For the validity on the modification due to the pion mass only in the TPE potential made above, 
we make an alternative chiral extrapolation directly in $1/a_0$ as $1/a_0 = c + d m_\pi^2$ using the current and previous lattice data, as shown in Fig.~\ref{fig:a0} (Right), even though data from different calculations possess different systematic uncertainties.
We obtain $1/a_0 = -0.01(9)$ fm$^{-1}$ at $m_\pi =135$ MeV, which is consistent with $1/a_0 = -0.03(4)$ fm$^{-1}$ with larger error.

  \subsection{$D^0D^0\pi^+$ mass spectrum}
  We construct $D^0D^0\pi^+$ mass spectrum by evaluating the $D^{*+}D^0$ rescattering   in the final state 
  based on the potential obtained in lattice QCD.  A production amplitude $U(M,p)$ with  invariant mass $M$
  and relative momentum $p$ for $D^{*+}D^0$ pair in the $I=0$ and $S$-wave channel is given by a direct generation from a constant interaction plus a rescattering (see the insect of Fig.~\ref{fig:ERE} (Right)  ) as
\begin{equation}
U(M,p) = P + \int {d^3\mathbf{q}^3\over (2\pi)^3} T(M,p,q) G(M,q) P,
\end{equation}
where the $T$-matrix $T(M,p,q)$ with in-coming (out-going) momentum $q$ ($p$) is obtained from the Lippmann-Schwinger equation with the lattice potential, and the  $D^{*+}D^0$ propagator $G(M,q)$ is given by $G^{-1}(M,q) = M - m_{D^{*+}} - m_{D^0} -{q^2\over 2\mu} + {i\over  2}\Gamma_{D^{*+}}$ with $\Gamma_{D^{*+}}=82.5$ keV being a value for the decay width of $D^{*+}$\cite{ParticleDataGroup:2022pth}. 
Then the $D^0D^0\pi^+$ mass spectrum is given by
\begin{equation}
{d {\rm Br} [D^0D^0\pi^+]\over dM} ={\cal N}  \int pdp \int \bar p d\bar p 
\left[ U(M,p) G(M,p) q_\pi(p) + U(M,\bar p) G(M,\bar p) q_\pi(\bar p) \right]^2, 
\end{equation}
where the pion momentum $q_\pi(p)$ comes from $D^{*+} \to D^0\pi^+$ decay and is determined kinematically,
and the second contribution arises from symmetrizing $D^0 D^0$ in the final state of the $D^0D^0\pi^+$ amplitude.
As the constant $P$ can be absorbed into an overall normalization ${\cal N}$, the shape of  $D^0 D^0 \pi^+$mass spectrum  is free from any unknown parameters.

Black and red bands in Fig.~\ref{fig:ERE} (Right)  are lattice predictions for  $D^0 D^0 \pi^+$mass spectrum at $m_\pi = 146.4$ MeV and $m_\pi = 135$ MeV, respectively, employing experimental values for $m_{D^{*+}}$, $m_{D^0}$ and $m_{\pi^+}$ in both cases to keep the same phase space with the experiment, while the black points represent LHCb data\cite{LHCb:2021vvq}.
While both lattice predictions have a peak around the $D^{*+} D^0$ threshold, 
the peak position moves to the left as $m_\pi$ decreases corresponding to the evolution from a  near threshold virtual pole to a loosely bound state pole
in the scattering amplitude.
The LHCb experimental data are better described by 
the red band at physical pion mass, whose peak appears just on the $D^{*+} D^0$ threshold, though visible differences still exist  above but near threshold region.

 \section{Summary}
We have employed the HAL QCD method to investigate a doubly charmed tetraquark state  $T_{cc}^+$ at nearly physical  pion mass $m_\pi= 146$ MeV. The $D^* D$ potential is attractive at all distances in the $I(J^P) = 0(1^+)$ channel. 
The system appears very close to unitarity, so that a small change in pion mass from 146 MeV to 135 MeV leads to significant changes in physical observables. As the pion mass decreases, the pole in the scattering amplitude evolves from a virtual pole to a bound state pole, and
the $D^0D^0\pi^+$ mass spectrum better reproduce the LHCb data. 

In this report,  the modification of the interaction due to the change of the pion mass is assumed to appear only in the exponent of the TPE potential
or the extrapolation of $1/a_0$  to $m_\pi=135$ MeV is performed linearly  in $m_\pi^2$ using data in the wide range of the pion mass. 
Even though two results are consistent, a direct calculation of scattering parameters,  which is feasible for the HAL QCD method 
but challenge for the finite volume method since an expected energy shift is very tiny,
is required for more reliable conclusions.
For this purpose,  we have just finished generating configurations in (2+1)-flavor QCD at $m_\pi \simeq 135$ MeV\cite{Itou:2023kvl}.
An inclusion of isospin breaking such as the $u,d$ quark mass difference and QED effect will be necessary in future but
extremely difficult since the coupled channel analysis  including 3-body states is required.

 \section*{Acknowledgements}
We thank members of the PACS Collaboration for the gauge configuration generation conducted on the K computer at RIKEN.
The lattice QCD measurements have been performed on Fugaku and HOKUSAI supercomputers at RIKEN.
We thank ILDG/JLDG((http://www.lqcd.org/ildg and http://www.jldg.org).
This work has been supported in part by HPCI System Research Project (hp120281, hp130023, hp140209, hp150223, hp150262, hp160211, hp170230, hp170170, hp180117, hp190103, hp200130, hp210165, hp210212, hp220240, hp220066, and hp220174),  the National Key R\&D Program of China (Contract No. 2018YFA0404400), the National Natural Science Foundation of China (Grant Nos. 11935003, 11975031, 11875075, 12070131001, and 12141501), the JSPS (Grant Nos. JP18H05236, JP22H00129, JP19K03879, JP18H05407, and JP21K03555),  ``Priority Issue on Post-K computer'' (Elucidation of the Fundamental Laws and Evolution of the Universe), ``Program for Promoting Researches on the Supercomputer Fugaku'' (Simulation for basic science: from fundamental laws of particles to creation of nuclei), and Joint Institute for Computational Fundamental Science (JICFuS).

\end{document}